
\font\gross=cmbx10  scaled\magstep2
\font\mittel=cmbx10 scaled\magstep1
\def\gsim{\mathrel{\raise.3ex\hbox{$>$\kern- .75em
                      \lower1ex\hbox{$\sim$}}}}
\def\lsim{\mathrel{\raise.3ex\hbox{$<$\kern-.75em
                      \lower1ex\hbox{$\sim$}}}}

\def\square{\kern 1pt
\vbox{\hrule height 0.6pt\hbox{\vrule width 0.6pt \hskip 3pt
\vbox{\vskip 6pt}\hskip 3pt\vrule width 0.6pt}
\hrule height 0.6pt}\kern 1pt }

\def\sla{\raise.15ex\hbox{$/$}\kern-.72em}


\parskip=\medskipamount
\overfullrule=0pt
\raggedbottom
\def\normalparindent{24pt}
\newif\ifdraft \draftfalse

\nopagenumbers
\footline={\ifnum\pageno=1 {\ifdraft
{\hfil\rm Draft \number\day -\number\month -\number\year}
\else{\hfil}\fi}
\else{\hfil\rm\folio\hfil}\fi}
\def\endpage{\vfill\eject}
\def\beginlinemode{\endmode\begingroup\parskip=0pt
\obeylines\def\\{\par}\def\endmode{\par\endgroup}}
\def\beginparmode{\endmode\begingroup \def\endmode{\par\endgroup}}
\let\endmode=\par
\def\raggedcenter{
              \leftskip=2em plus 6em \rightskip=\leftskip
             \parindent=0pt \parfillskip=0pt \spaceskip=.3333em
            \xspaceskip=.5em\pretolerance=9999 \tolerance=9999
              \hyphenpenalty=9999 \exhyphenpenalty=9999 }
\def\\{\cr}
\let\rawfootnote=\footnote
\def\footnote#1#2{{\parindent=0pt\parskip=0pt
\rawfootnote{#1}{#2\hfill\vrule height 0pt depth 6pt width 0pt}}}
\def\title
{\null\vskip 3pt plus 0.2fill\beginlinemode\raggedcenter\gross}
\def\author{\vskip 3pt plus 0.2fill \beginlinemode\raggedcenter}
\def\affil{\vskip 3pt plus 0.1fill\beginlinemode\raggedcenter\it}
\def\abstract{\vskip 3pt plus 0.3fill \beginparmode{\noindent
{\mittel Abstract}:~}  }
\def\endtitlepage{\endpage\body}
\def\body{\beginparmode\parindent=\normalparindent}
\def\head#1{\par\goodbreak{\immediate\write16{#1}
\vskip 0.4cm{\noindent\gross
#1}\par}\nobreak\nobreak\nobreak\nobreak}

\def\finalcite{\citeall\ref\citeall\Ref}
\newif\ifannpstyle
\newif\ifprdstyle
\newif\ifplbstyle
\newif\ifwsstyle

\gdef\refto#1{\ifprdstyle  $^{\[#1] }$ \else
              \ifwsstyle$^{\[#1]}$  \else
              \ifannpstyle $~[\[#1] ]$ \else
              \ifplbstyle  $~[\[#1] ]$ \else
                                         $^{[\[#1] ]}$\fi\fi\fi\fi}
\gdef\refis#1{\ifprdstyle \item{~$^{#1}$}\else
              \ifwsstyle \item{#1.} \else
              \ifplbstyle\item{~[#1]} \else
              \ifannpstyle \item{#1.} \else
                              \item{#1.\ }\fi\fi\fi\fi }
\gdef\journal#1,#2,#3,#4.{
           \ifprdstyle {#1~}{\bf #2}, #3 (#4).\else
           \ifwsstyle {\it #1~}{\bf #2~} (#4) #3.\else
           \ifplbstyle {#1~}{#2~} (#4) #3.\else
           \ifannpstyle {\sl #1~}{\bf #2~} (#4), #3.\else
                       {\sl #1~}{\bf #2}, #3 (#4)\fi\fi\fi\fi}

\def\ref#1{Ref.~#1}
\def\Ref#1{Ref.~#1}
\def\cite#1{{#1}}\def\[#1]{\cite{#1}}

\def\eq#1{Eq.~\(#1)}\def\eqs#1{Eqs.~\(#1)}

\def\(#1){(\call{#1})}
\def\call#1{{#1}}\def\taghead#1{{#1}}

\def\references{\head{References}
\beginparmode\frenchspacing\parskip=0pt}
\def\endreferences{\body}
\def\endit{\endmode\vfill\supereject}\let\endpaper=\endit


\def\kndir{Universit\"at Konstanz, Fakult\"at f\"ur Physik, Postfach
5560,
D-7750, Konstanz, Germany.\\ e-mail: phlousto@dknkurz1.}

\def\iafedir{Permanent Address: IAFE, Cas. Corr. 67, Suc. 28,
1428 Buenos Aires, ARGENTINA.\\ e-mail: lousto@iafe.edu.ar.}

\def\mangos{This work was partially supported
by the Directorate General for
Science, Research and Development of
the Commission of the European Communities
and by the Alexander von Humboldt Foundation.}

\catcode`@=11
\newcount\r@fcount \r@fcount=0\newcount\r@fcurr
\immediate\newwrite\reffile\newif\ifr@ffile\r@ffilefalse
\def\w@rnwrite#1{\ifr@ffile\immediate\write\reffile{#1}\fi\message{#1
}}
\def\writer@f#1>>{}
\def\referencefile{\r@ffiletrue\immediate\openout\reffile=\jobname.re
f%
  \def\writer@f##1>>{\ifr@ffile\immediate\write\reffile%
    {\noexpand\refis{##1} = \csname r@fnum##1\endcsname = %
     \expandafter\expandafter\expandafter\strip@t\expandafter%
\meaning\csname r@ftext\csname r@fnum##1\endcsname\endcsname}\fi}%
  \def\strip@t##1>>{}}

\def\citeall#1{\xdef#1##1{#1{\noexpand\cite{##1}}}}
\def\cite#1{\each@rg\citer@nge{#1}}
\def\each@rg#1#2{{\let\thecsname=#1\expandafter\first@rg#2,\end,}}
\def\first@rg#1,{\thecsname{#1}\apply@rg}
\def\apply@rg#1,{\ifx\end#1\let\next=\relax%
\else,\thecsname{#1}\let\next=\apply@rg\fi\next}%
\def\citer@nge#1{\citedor@nge#1-\end-}
\def\citer@ngeat#1\end-{#1}
\def\citedor@nge#1-#2-{\ifx\end#2\r@featspace#1
  \else\citel@@p{#1}{#2}\citer@ngeat\fi}
\def\citel@@p#1#2{\ifnum#1>#2{\errmessage
{Reference range #1-#2\space is bad.}
\errhelp{If you cite a series of references
by the notation M-N, then M and
N must be integers, and N must be greater than or equal to M.}}\else%
{\count0=#1\count1=#2\advance\count1
by1\relax\expandafter\r@fcite\the\count0,%
  \loop\advance\count0 by1\relax
    \ifnum\count0<\count1,\expandafter\r@fcite\the\count0,%
  \repeat}\fi}
\def\r@featspace#1#2 {\r@fcite#1#2,}
\def\r@fcite#1,{\ifuncit@d{#1}
    \expandafter\gdef\csname r@ftext\number\r@fcount\endcsname%
{\message{Reference #1 to be supplied.}
\writer@f#1>>#1 to be supplied.\par
     }\fi\csname r@fnum#1\endcsname}
\def\ifuncit@d#1{\expandafter\ifx\csname r@fnum#1\endcsname\relax%
\global\advance\r@fcount by1%
\expandafter\xdef\csname r@fnum#1\endcsname{\number\r@fcount}}
\let\r@fis=\refis   \def\refis#1#2#3\par{\ifuncit@d{#1}%
\w@rnwrite{Reference #1=\number\r@fcount
\space is not cited up to now.}\fi%
  \expandafter\gdef\csname r@ftext\csname
  r@fnum#1\endcsname\endcsname%
  {\writer@f#1>>#2#3\par}}
\def\r@ferr{\endreferences\errmessage{I was expecting to see
\noexpand\endreferences before now;  I have inserted it here.}}
\let\r@ferences=\references
\def\references{\r@ferences\def\endmode{\r@ferr\par\endgroup}}
\let\endr@ferences=\endreferences
\def\endreferences{\r@fcurr=0{\loop\ifnum\r@fcurr<\r@fcount
\advance\r@fcurr by 1\relax\expandafter\r@fis
\expandafter{\number\r@fcurr}%
    \csname r@ftext\number\r@fcurr\endcsname%
  \repeat}\gdef\r@ferr{}\endr@ferences}
\let\r@fend=\endpaper\gdef\endpaper{\ifr@ffile
\immediate\write16{Cross References written on
[]\jobname.REF.}\fi\r@fend}
\catcode`@=12
\finalcite
\catcode`@=11
\newcount\tagnumber\tagnumber=0
\immediate\newwrite\eqnfile\newif\if@qnfile\@qnfilefalse
\def\write@qn#1{}\def\writenew@qn#1{}
\def\w@rnwrite#1{\write@qn{#1}\message{#1}}
\def\@rrwrite#1{\write@qn{#1}\errmessage{#1}}
\def\taghead#1{\gdef\t@ghead{#1}\global\tagnumber=0}
\def\t@ghead{}\expandafter\def\csname @qnnum-3\endcsname
  {{\t@ghead\advance\tagnumber by -3\relax\number\tagnumber}}
\expandafter\def\csname @qnnum-2\endcsname
  {{\t@ghead\advance\tagnumber by -2\relax\number\tagnumber}}
\expandafter\def\csname @qnnum-1\endcsname
  {{\t@ghead\advance\tagnumber by -1\relax\number\tagnumber}}
\expandafter\def\csname @qnnum0\endcsname
  {\t@ghead\number\tagnumber}
\expandafter\def\csname @qnnum+1\endcsname
  {{\t@ghead\advance\tagnumber by 1\relax\number\tagnumber}}
\expandafter\def\csname @qnnum+2\endcsname
  {{\t@ghead\advance\tagnumber by 2\relax\number\tagnumber}}
\expandafter\def\csname @qnnum+3\endcsname
  {{\t@ghead\advance\tagnumber by 3\relax\number\tagnumber}}
\def\equationfile{\@qnfiletrue\immediate
\openout\eqnfile=\jobname.eqn%
  \def\write@qn##1{\if@qnfile\immediate\write\eqnfile{##1}\fi}
  \def\writenew@qn##1{\if@qnfile\immediate\write\eqnfile
    {\noexpand\tag{##1} = (\t@ghead\number\tagnumber)}\fi}}
\def\callall#1{\xdef#1##1{#1{\noexpand\call{##1}}}}
\def\call#1{\each@rg\callr@nge{#1}}
\def\each@rg#1#2{{\let\thecsname=#1\expandafter\first@rg#2,\end,}}
\def\first@rg#1,{\thecsname{#1}\apply@rg}
\def\apply@rg#1,{\ifx\end#1\let\next=\relax%
\else,\thecsname{#1}\let\next=\apply@rg\fi\next}
\def\callr@nge#1{\calldor@nge#1-\end-}\def\callr@ngeat#1\end-{#1}
\def\calldor@nge#1-#2-{\ifx\end#2\@qneatspace#1 %
  \else\calll@@p{#1}{#2}\callr@ngeat\fi}
\def\calll@@p#1#2{\ifnum#1>#2{\@rrwrite{Equation range #1-#2\space is
bad.}
\errhelp{If you call a series of equations by the notation M-N, then
M and
N must be integers, and N must be greater than or equal to M.}}\else%
{\count0=#1\count1=#2\advance\count1 by1
\relax\expandafter\@qncall\the\count0,%
  \loop\advance\count0 by1\relax%
 \ifnum\count0<\count1,\expandafter\@qncall\the\count0,  \repeat}\fi}
\def\@qneatspace#1#2 {\@qncall#1#2,}
\def\@qncall#1,{\ifunc@lled{#1}{\def\next{#1}\ifx\next\empty\else
\w@rnwrite{Equation number \noexpand\(>>#1<<) has not been defined
yet.}
 >>#1<<\fi}\else\csname @qnnum#1\endcsname\fi}
\let\eqnono=\eqno\def\eqno(#1){\tag#1}
\def\tag#1$${\eqnono(\displayt@g#1 )$$}
\def\aligntag#1\endaligntag  $$
{\gdef\tag##1\\{&(##1 )\cr}\eqalignno{#1\\}$$
  \gdef\tag##1$${\eqnono(\displayt@g##1 )$$}}
  
\def\eqalignno#1{\displ@y \tabskip\centering
  \halign to\displaywidth{\hfil$\displaystyle{##}$\tabskip\z@skip
    &$\displaystyle{{}##}$\hfil\tabskip\centering
    &\llap{$\displayt@gpar##$}\tabskip\z@skip\crcr
    #1\crcr}}
\def\displayt@gpar(#1){(\displayt@g#1 )}
\def\displayt@g#1 {\rm\ifunc@lled{#1}\global\advance\tagnumber by1
        {\def\next{#1}\ifx\next\empty\else\expandafter
   \xdef\csname @qnnum#1\endcsname{\t@ghead\number\tagnumber}\fi}%
  \writenew@qn{#1}\t@ghead\number\tagnumber\else
        {\edef\next{\t@ghead\number\tagnumber}%
        \expandafter\ifx\csname @qnnum#1\endcsname\next\else
 \w@rnwrite{Equation \noexpand\tag{#1} is a duplicate number.}\fi}%
  \csname @qnnum#1\endcsname\fi}
\def\eqnoa(#1){\global\advance\tagnumber by1\multitag{#1}{a}}
\def\eqnob(#1){\multitag{#1}{b}}
\def\eqnoc(#1){\multitag{#1}{c}}
\def\eqnod(#1){\multitag{#1}{d}}
\def\multitag#1#2$${\eqnono(\multidisplayt@g{#1}{#2} )$$}
\def\multidisplayt@g#1#2 {\rm\ifunc@lled{#1}
        {\def\next{#1}\ifx\next\empty\else\expandafter
    \xdef\csname @qnnum#1\endcsname{\t@ghead\number\tagnumber b}\fi}%
  \writenew@qn{#1}\t@ghead\number\tagnumber #2\else
        {\edef\next{\t@ghead\number\tagnumber #2}%
 \expandafter\ifx\csname @qnnum#1\endcsname\next\else
\w@rnwrite{Equation \noexpand\multitag{#1}{#2} is a duplicate
number.}
\fi}%
  \csname @qnnum#1\endcsname\fi}
\def\ifunc@lled#1{\expandafter\ifx\csname @qnnum#1\endcsname\relax}
\let\@qnend=\end\gdef\end{\if@qnfile
\immediate\write16{Equation numbers written on []\jobname.EQN.}
\fi\@qnend}

\magnification=1200
\baselineskip=12pt
\def\pnt{\par\noindent}
\def\k{\kappa}
\title

The Fourth Law of Black Hole Thermodynamics

\author
C. O. Lousto\footnote{$^*$}{\iafedir}

\affil\kndir

\abstract

We show that black holes fulfill the scaling laws
arising in critical transitions. In particular, we find that in
the transition from negative to positive values the heat capacities
$C_{JQ}$, $C_{\Omega Q}$ and $C_{J\Phi}$ give rise to
critical exponents satisfying the scaling laws. The three transitions
have the same critical exponents as predicted by the universality
Hypothesis. We also briefly
discuss the implications of this result with regards to the
connections
among gravitation, quantum mechanics and statistical physics.

\bigskip\bigskip\bigskip


\endtitlepage
\baselineskip=12pt
\head{1. The four laws of Black Hole Thermodynamics}

In the work of Bardeen, Carter and Hawking \refto{1} it was
established a
remarkable mathematical analogy between the laws of  thermodynamics
and the
laws of black hole mechanics derived from General Relativity. If one
makes
the formal replacements $E\to M$, $T\to C\k$, and  $S\to A/8\pi C$
(where
$C$ is a constant) in the laws of the thermodynamics, one obtains the
laws
that govern the mechanics of black holes \refto{9}. The physical
analogy
seemed to have problems due  to the fact that in classical General
Relativity
the thermodynamic temperature of a black hole appears to be absolute
zero.
However, Hawking \refto{2} found that when quantum effects are taken
into
account, a black hole absorbs and emits particles as a body
at temperature $T=\k/2\pi$, and this resolved that puzzle. Here and
throughout this paper we take units in which $G=\hbar=c=\k_B=1$.

The four laws of black hole thermodynamics can be briefly formulated
as follows:

\bigskip
{\it The Zeroth Law:} The surface  gravity, $\k$, of
a stationary black hole (at
equilibrium) is constant on the entire surface of the event horizon.

This property is proved as a theorem in ref. [\cite{1}].
Landsberg \refto{6} gives however a more precise formulation: ``In
the absence
of adiabatic partitions and long range fields, an equilibrium system
exhibits
a unique temperature". If the above conditions are not satisfied, it
can be
shown \refto{7} that a system with a built-in adiabatic partition can
have
two  different temperatures even in thermal equilibrium.

For a Kerr-Newman black hole endowed with mass $M$, charge $Q$ and
angular
momentum $\vec J$, the surface gravity is given by \refto{3}
$$
\k={1\over 2} {r_+ - r_-\over (r_+^2+a^2)}={\sqrt{M^2-a^2-Q^2}\over
2M^2-Q^2+2M\sqrt{M^2-a^2-Q^2}}~~, \eqno(1)
$$
where $a^2=|\vec J|^2/M^2$ and
$$
r_{\pm}=M\pm\sqrt{M^2-a^2-Q^2}~~, \eqno(2)
$$
are the event and internal horizons respectively.

\bigskip
{\it The First Law:} This is just an expression that states that in
an isolated
system, including black holes, the total energy of the system is
conserved.
In ref.[\cite{1}] it is derived, in a general context, the following
differential mass formula for stationary black holes,
$$
\delta M= {\k\over 8\pi}\delta A+ \vec\Omega\cdot\delta\vec J
+\Phi\delta
Q~~, \eqno(3)
$$
where $\vec\Omega$ is the angular velocity and $\Phi$ the electric
potential
of the event horizon, and
$$
A=4\pi(r_+^2+a^2)=8\pi\left\{M^2-{Q^2\over
2}+\sqrt{M^4-J^2-M^2Q^2}\right\}~~,
 \eqno(4)
$$
is its area. $Mc^2$ represents actually the total energy of the hole,
$E$,
and $T=\k/2\pi$ is its Bekenstein-Hawking temperature.

Eq. \(3) shows that $A/4$ is the analogous of the entropy of the
black hole
$$
S_{bh}={1\over 4} A~~. \eqno(5)
$$
The second and third term on the right hand side of Eq.\(3) represent
the work
done or energy extracted when we change the black hole angular
momentum and
electric charge respectively. The angular velocity and electric
potential are
given by \refto{5}
$$
\Omega={a\over r_+^2+a^2}~~, \eqno(6)
$$
$$
\Phi={Q r_+ \over  r_+^2 + a^2}~~. \eqno(7)
$$

Inverting relation \(4), one obtains\refto{ch,cr}
the mass as a function of $A$, $J$ and $Q$
$$
M=\left({A\over 16\pi}+{4\pi J^2\over A}+{Q^2\over 2}+{\pi Q^4\over
A}\right)^{1/2}~~. \eqno(8)
$$
This is the fundamental thermodynamic relation containing explicitly
all the
information about the thermodynamic state of the black hole.

By applying Euler's theorem on homogeneous functions to $M$, which is
homogeneous of degree $1/2$, one obtains \refto{5}
$$
M={1\over 2}TA+2\vec\Omega\cdot\vec J+\Phi Q~~. \eqno(9)
$$
Thus, $T$, $\Omega$ and $\phi$ are the ``intensive" parameters and
are constant
everywhere on the event horizon of any stationary axisymmetric black
hole.

\bigskip
{\it The Second Law:} Bekenstein \refto{9,45} proposed that the total
entropy
(see also \refto{10})
$$
S=S_{bh}+S_{m}~~, \eqno(10)
$$
(where $S_{m}$ is the total entropy of ordinary matter outside black
holes)
never decreases in any physical process.

Hawking \refto{11} probed a theorem stating that the area $A$ of the
event
horizon of each black hole can not decrease with time, i.e.
$$
\delta A\geq 0 ~~. \eqno(11)
$$
However, when quantum processes close to the event horizon are taken
into
account, this  theorem can be violated since the $T_{\mu\nu}$ of the
emitted
Hawking radiation \refto{2} does not respect the positive energy
condition
assumed in the proof of this theorem. In fact, an isolated black hole
can
eventually evaporate completely, thus decreasing its area to zero.
On the other hand, one can easily make decrease the total entropy of
matter
outside black holes, $S_m$, by letting fall that matter into a black
hole.

When both addends on the right hand side of eq.\(10) are taken into
account,
a decrease in one of them seems to be always compensated by an
increase in
the other.
In fact, when matter is thrown into a black hole, thus decreasing
$S_m$, the
area of the black hole will tend to increase. Conversely, when a
black hole
evaporates  emitting particles, the matter outside the black hole is
in a
higher
entropy state. Thus we have  for any process
$$
\delta S\geq 0~~. \eqno(12)
$$
This means that in any dynamical process the system will tend towards
an
equilibrium state for which the entropy has the largest possible
value
subject
to the given structure of the system \refto{6}.
The ordinary second law for $S_m$ would be a special case of eq.\(12)
applicable
when black holes are not present, whereas the area theorem, eq.\(11),
would
apply in the classical limit where the ordinary entropy flux into a
black hole
is positive.

The validity of eq.\(12) was challenged by several gedanken
processes,
although
when semiclassical corrections are property taken into account,
no violation of the second law can be achieved \refto{12}.
This strongly suggest that eq\(12)
must hold, at least for quasi stationary processes when departures
from
equilibrium are small \refto{8}.

\bigskip
{\it The Third Law:} The usual formulation of the third law states
that
\refto{1}
it is impossible by any physical process to reduce $\k$ to zero by a
finite
sequence of operations.

This is the analogue of the third law of thermodynamics as formulated
by Nerst.
Note that Plank's formulation stating that the entropy of any system
tends to
an absolute constant, (which may be taken as zero), when  $T\to 0$,
does not
hold in black holes dynamics. In fact, for an extreme Kerr-Newman
black hole,
$$
S_{bh}(T=0)=\pi(Q^2+2a^2)~~. \eqno(13)
$$
Page \refto{page} has conjectured that it should actually be
$S(T=0)=0$, for
the ground state of a black hole not to be degenerate. However,
let us  note that $\k=0$ can be also reached (even for Schwarzschild
black
holes)
in the limit $M\to \infty$ (as can  be seen from eqs. \(1)-\(2))
while $S\to
\infty$ in this case.

Israel \refto{13a,13b} has given a formulation and proof of the third
law of
black hole dynamics for Reissner-Nordstr\o m holes. It can be stated
informally
as follows:

``A non extremal black hole cannot become extremal (i. e., lose its
trapped
surfaces) within a finite interval of advanced time in any continuous
process
in which the stress-energy tensor of infalling matter remains bounded
and
satisfies the weak energy conditions in a neighborhood of the outer
apparent
horizon".
For a more formal statement of this formulation see refs.
[\cite{13a,13b}].

There is a close relation between the validity of the third law and
the cosmic
censorship hypothesis. In fact, Israel's Lemma can be used to
stablish a weak
form of this hypothesis called ``gravitational confinement"
\refto{13b}.
If one would reduce the value of $\k$  of a Kerr-Newman black hole by
throwing
in particles to increase the angular momentum and/or the electric
charge, up
to the state where $a^2+Q^2=M^2$ in a finite number of steps (it is
not possible
by mining \refto{Roman}), then presumably one could carry the process
further
up to $a^2+Q^2>M^2$. The result would be a naked singularity. Then
the
singularity would no longer be hidden  inside a black hole but would
be able
to influence, and be observed by, the outside universe. This is so
unpleasant
that one invokes the cosmic censorship conjecture, thus ensuring the
validity
of the third law. This theorem, however, remains elusive to any
proof, being
one of the most important unsolved problems in classical General
Relativity.

In the next section we will review some important results concerning
phase
transitions. These results can be resumed by the scaling laws of
thermodynamics. Then, in the third section we postulate
this laws to hold for black holes and study
its validity in the case of the black hole phase transitions found by
Davies
\refto{4, 14} and its extensions to other heat capacities\refto{26}.
We end the paper with some discussion on this results and a
future  prospect of research on the subject.

\head{2. Scaling  in critical phenomena}

The existence of critical phase transitions was discovered by Andrews
\refto{15}
 in his studies of carbon dioxide. Liquid and vapor phases became
identical at
 the critical point and there was a seemingly continuous transition
from one
 phase to the other. The phase equilibrium curve in the PT-plane
terminate at
 a certain point, called critical point. The corresponding
temperature and
  pressure are the critical temperature, $T_c$,  and the critical
pressure,
  $P_c$. At this point, liquid and vapor become indistinguishable. By
going
  round the critical point, it is possible to take a path from the
liquid
  region to the vapor region without crossing any phase boundary and
thus
  without experiencing any discontinuous change in properties. This
suggest
   that  critical points can only exist for phases such that the
difference
  between them  is purely quantitative (for example liquid-gas) and
not
  qualitative such the case of a solid (crystal), since they have
different
  internal symmetry.

According to the thermodynamic inequality,
${\delta P/\delta V}\big\vert_{T}<0 ~,$
$P$ is a decreasing function of $V$. However, in the critical state
$$
{\delta P\over\delta V}\biggr\vert_{T}(c)=0~~,~~
{\delta^2 P\over\delta V^2}\biggr\vert_{T}(c)=0~~.\eqno(14)
$$

{}From the formula
$$
C_{P}-C_{V}=-T{\left({\delta P\over\delta
T}\big\vert_{V}\right)^2\over
\left({\delta P\over\delta V}\right)_T}  ~~, \eqno(16)
$$
for the difference of specific heats,
we conclude that $C_{P}\to\infty$ at the critical point (For more
details
 see, for example, ref.[\cite{16}]).

In the magnetic case a continuous phase transition occurs from an
ordered
 ferromagnetic state to a paramagnetic state. The critical point is
at zero
applied magnetic field $\vec H$ and at $T=T_c$. The derivative of the
magnetization, $\vec M$, diverge at $T_c$. The phase transition has
no
associated
 latent heat and can be described as a critical phase transition. An
important
 difference arises with respect to the liquid-gas system because the
analogue
 of $P$ and  $V$ variables,  i. e., $\vec H$ and  $\vec M$,  are
vector
 quantities. However, in an ideal magnet, we neglect the anisotropies
and the
 magnetic properties will only depend on the magnitude of  $\vec H$.

The same general properties of critical phase transitions are also
found for
 ferroelectricity, superconductivity, superfluidity in liquid helium,
mixing
 of liquids, and ordering in alloys (see ref.\refto{17} for further
details).
  Wilson \refto{18} also mentioned turbulent fluid flow, internal
structure of
   elementary particles and the interaction between electrons in a
metal with
  magnetic impurities.

\bigskip
To give a precise and general definition of  {\it critical point
exponents}
in describing the behavior near the critical point of a general
function
$f(x)$ we assume that the following limit exists
$$
\sigma \doteq \lim_{\varepsilon\to 0} {\ln {f(\varepsilon)}\over \ln
{\varepsilon}}~~,~~\varepsilon\doteq T-T_c~~, \eqno(17)
$$
where $\sigma$ is the critical point exponent of $f(\varepsilon)$.

Away from the critical point we will have deviations of the form
\refto{30}
$$
f(\varepsilon)=A \varepsilon^{\sigma}(1+B\varepsilon^y+...)~~,
{}~~~y>0~~.
$$
For magnetic systems, in particular, it is possible to define the
following
thermodynamical critical point exponents as $\varepsilon\to 0^+$:\pnt
For the specific heat at constant magnetic field (note that here one
departs
from the strict fluid-magnet analogy $V\leftrightarrow -M$ and
$P\leftrightarrow
H$( see \refto{19})):
$$
C_H=T {\partial S\over\partial T}\biggr\vert_{H}=\cases
{\sim \varepsilon^{-\alpha},  &for $H=0$ \cr
\sim H^{\varphi}, &for $\varepsilon=0$ \cr} ~~, \eqno(19)
$$
zero- field magnetization:
$$
M=\cases
{\sim \varepsilon^{\beta},  &for $H=0$ \cr
\sim H^{\delta ^{-1}}, &for $\varepsilon=0$ \cr} ~~, \eqno(20)
$$
zero-field isothermal susceptibility.
$$
\chi_{_{T}}={\partial M\over\partial H}\biggr\vert_{T}=\cases
{\sim \varepsilon^{-\gamma},  &for $H=0$ \cr
\sim H^{1-\delta ^{-1}}, &for $\varepsilon=0$ \cr}   ~~, \eqno(21)
$$
and critical  entropy
$$
 S-S_c=\cases
{\sim \varepsilon^{ 1-\alpha},  &for $H=0$ \cr
\sim H^{\psi}, &for $\varepsilon=0$ ~~.\cr} \eqno(22)
$$
When we are close to, but just below the critical temperature (at
zero field),
 we replace $\varepsilon$ by $-\varepsilon$ and the critical
exponents by  its
  primed analogs $\alpha'$, $\beta'$, $\gamma'$ and $\delta'$ (For
more details
 and notation see, for example, ref [\cite{19}]).

\bigskip
We now state the scale invariance hypothesis or {\it fourth law of
thermodynamics},
 \refto{19, 20, 21}:\pnt
 {\it Close to the critical point the singular part of the Gibbs free
energy is
  a generalized homogeneous function of its variables}. Thus we have,
 $$
 G(\lambda^{a_{\varepsilon}}\varepsilon, \lambda^{a_{H}}H)=\lambda
 G(\varepsilon, H),
 \eqno(23)
 $$
 where $a_{\varepsilon}$ and $a_{H}$ are two parameters called
scaling powers,
  and $\lambda$ is arbitrary.

 The scaling hypothesis for static critical phenomena have been made
\refto{22}
  in a variety of situations and applied to thermodynamics functions
and to
  static and dynamics correlation functions.

 The above hypothesis transcends an ad hoc assumption to deal with
the behavior
of thermodynamic functions near the critical point. Its importance,
tightly
related to the presence of a symmetry (which is spontaneously broken)
underlying the system,
transforms this hypothesis in a true
law of thermodynamics associated to the invariance of the behavior of
a
system with respect to scale transformations  when it is under
critical
conditions.

 It is clear that once accepted the condition of generalized function
for one
 of the thermodynamic potentials, the same property holds for all the
other
 thermodynamic potentials, since the Legendre transformations that
relate each
 other conserve the mathematical content of the functions they are
applied to
 \refto{19, 20}.

 Using eq.\(23),  all the critical point exponents can be simply
expressed in
  terms of $a_{\varepsilon}$ and $a_{H}$. In fact, since
$$
C_H=T {\partial^2 G\over\partial  T^2}\biggr\vert_{H}~~;~~~
M=-{\partial
 G\over\partial  H}\biggr\vert_{T}~~;~~S=-{\partial  G\over\partial
T}\biggr\vert_{H}~~~~\rm{and}~~~\chi_{_{T}}=-{\partial^2
G\over\partial
  H^2}\biggr\vert_{T}~~,
\eqno(24')
$$
we can relate the various critical exponents to the two scaling
parameters
 $a_{\varepsilon}$ and $a_{H}$. Thus, one obtains \refto{19,20}
$$
\alpha=2-{1\over a_{\varepsilon}}~~; ~~\beta={1-a_{H}\over
 a_{\varepsilon}}~~;~~\delta={ a_{H}\over 1-a_{H}}~~;$$
$$\gamma={2a_{H}-1\over
a_{\varepsilon}}~~;~~\psi={1-a_{\varepsilon}\over
a_{H}}~~;~~\varphi={1-2a_{\varepsilon}\over a_{H}}~~. \eqno(24'')
$$
  We can further eliminate this two parameters and obtain a set of
equalities
   among the exponents called {\it scaling laws}: \refto{19}
$$
\alpha+2\beta+\gamma=2~~,
$$
$$
\alpha+\beta(\delta +1)=2 ~~,
$$
$$
\gamma(\delta +1)=(2-\alpha)(\delta-1)~~,
$$
$$
\gamma=\beta(\delta -1)~~, \eqno(24)
$$
$$
(2-\alpha)(\delta\psi-1)+1=(1-\alpha)\delta~~,
$$
$$
\varphi+2\psi-\delta^{-1}=1~~.
$$
These relations are not all independent of one another. Some of this
equalities were predicted as inequalities under stability
considerations. In
fact the first and second to fourth equations, fulfilled as
inequalities in
\(24), are known as the Rushbrooke and Griffiths inequalities
respectively.

A second important result of the static scaling is the equality of
primed and
unprimed critical point exponents.

Experimentally and theoretically it was observed that the critical
exponents
are rather insensitive to the details of the system. This observation
is
embodied in the {\it universality hypothesis} that states that for a
continuous
phase transition the static critical exponents depend only on the
following
three properties: \pnt
1) the dimensionality of the system, $d$.\pnt
2) the internal symmetry dimensionality of the order parameters,
$D$.\pnt
3) whether the forces are of short or long range.

The renormalization group approach \refto{18, 23} use the scaling
hypothesis
and provides a sound mathematical foundation to the concept of
universality.

The scaling is found to hold (within experimental error) in almost
every case.
 The renormalization group approach set the scaling theory in a
broader context
  and explains the circumstances under which it can  and how it
breaks down
  \refto{24, 25}.

\head{3. The Scaling Hypothesis applied to black holes}

As we have recalled in the last section, the scaling of critical
phenomena
applies to a great variety of thermodynamical systems. Those ranging
from
the internal structure of elementary particles to ferroelectricity
and
turbulent fluid flow, passing through superconductivity and
superfluidity.
We have also seen that black hole dynamics is governed by analogues
of
the ordinary four laws of thermodynamics and, with the appropriate
cares
corresponding to self - gravitating systems (negative heat
capacities), one
can apply these laws of
thermodynamics to any particular process underwent by a black hole.
This
two facts lead us to conjecture that black holes also obey the
scaling
laws or fourth law of thermodynamics:

{\it In the neighborhood of a critical point the singular part of
Helmholtz
 free energy,
$F(T,\vec J,Q)$, is a generalized homogeneous function of its
variables}.

Here we have established the hypothesis in terms of the Helmholtz
potential,
$F=M-TS$, for practical reasons that will become clear later, but
they are
essentially that we can write
$$
C_{JQ}=-T{\partial^2F\over\partial T^2}\biggr\vert_{J,Q}~,\eqno(25')
$$
and that this heat capacity reveals the critical transition suffered
by Kerr -
Newman black holes\refto{4}.

It can be seen from eq \(8) that $M(S,J,Q)$, corresponding in this
case
to the total energy of the black hole, is a homogeneous function of
degree one
 half. However, it will turn out, as we will see soon, that close to
the
 critical
transition the thermodynamic potential will become a linear
homogeneous
function of its variables measuring the departure from the critical
values.
Thus allowing the normal relations between thermodynamic functions to
be valid also in black hole thermodynamics. In particular, the
entropy
will recover (only in the neighborhood of the critical point) its
property
of {\it extensitivity} (this property  was, in turn, called by
Landsberg
\refto{28,29} the fourth law).

We  will next study the behavior of the black hole variables as it
undergoes
the phase transition first discovered by Davies\refto{4,14}. Thus, we
will
have the opportunity to  explicitly  check if black holes fulfill the
scaling laws, i.e., eqs \(24).

\bigskip

Let us suppose that a rotating charged black hole is held in
equilibrium at
some temperature $T$, with a surrounding heat bath. If we consider a
small,
reversible transfer of energy between the hole and its environment;
this
absorption will be isotropic, and will occur in such a way that the
angular
momentum $J$ and charge $Q$ remain unchanged, on the average. The
full
thermal capacity (not per unit mass) corresponding to this energy
transfer can
be computed by eliminating $M$ between eqs \(1) and \(8), and
differentiate
keeping $J$ and $Q$ constant,
$$C_{J,Q}=T{\partial S\over\partial
T}\biggr\vert_{J,Q}={MTS^3\over\pi J^2+
{\pi\over4}Q^4-T^2S^3}~.\eqno(25)$$

This heat capacity goes from negative values for a Schwarzschild
black hole,
$C_{Sch}=-M/T$, to positive values for a nearly extreme Kerr - Newman
black
hole, $C_{EKN}\sim \sqrt{M^4-J^2-M^2Q^2} \to 0^+$. Thus, $C_{J,Q}$
has changed
sign at some value of
$J$ and $Q$ in between. In fact, the heat capacity passes from
negative to
positive values through an infinite discontinuity. This feature has
lead
Davies\refto{4} to classify the phenomenon at the critical values of
$J$ and $Q$ as a second order phase transition. The values $J_c$ and
$Q_c$
at which the transition occurs are obtained by making to vanish the
denominator
on the right hand side of eq \(25). We can then define the following
parametrization,
$$
J_c^2={j\over8\pi}M^4~~~{\rm and}~~~Q_c^2={q\over8\pi}M^2~.\eqno(26)
$$
Eliminating $S$ and $T$ in eq \(25) by use of eqs \(1) and \(8),
the infinite discontinuity in $C_{J,Q}$ takes place at\refto{4}
$$
j_{JQ}^2+6j_{JQ}+4q_{JQ}=3~.\eqno(27)
$$
For an uncharged, i.e., Kerr, hole, $q_{JQ}=0$. Thus,
$j_{JQ}=2\sqrt{3}-3$.
Then we have
$$
\Omega_c={\sqrt{2\sqrt{3}-3}\over4\sqrt{3}-3}T_c\cong0.233T_c~.
\eqno(28)
$$
While for a non rotating, i.e., Reissner - Nordst\o m, hole,
$j_{JQ}=0$. Thus,
 $q_{JQ}=3/4$. And the critical value of the electric potential is
given by
$$
\Phi_c={1\over\sqrt{3}}~,\eqno(29)
$$
independent of the other parameters of the black hole such as its
mass or charge.

In ref [\cite{14}] Davies has extended this analysis of the heat
capacity
discontinuity to the case of  a Kerr - Newman black hole embedded in
a
De Sitter space. Hut \refto{32}  has also considered the effect of
the
surrounding radiation on the heat capacity  $C_v$  and found
explicitly the
existence of a critical point, it being the terminal point of a
phaseline.

It can also be shown\refto{26} that
the four isothermal compressibilities are divergent as their
corresponding
heat capacities (however, adiabatic compressibilities are non -
singular).
For example,
$$
K_{T,Q}^{-1}=J{\partial\Omega\over\partial J}\biggr\vert_{T,Q}\sim
{\pi(2\Phi Q-M)(1-4\pi TM)\over S^2[1-12\pi
TM+4\pi^2T^2(6M^2+Q^2)]}~,
\eqno(30)
$$
diverges as $C_{J,Q}$.
Also
$K_{T,J}^{-1}=C_{J,Q}(\partial\Phi/\partial
Q\big\vert_{S,J})/C_{J,\Phi}$
diverges as $C_{J,Q}$ on the singular
segment given by eqs \(26) - \(27).

By use of eqs \(8) and \(1),  the heat
capacity $C_{JQ}$ can be expressed as \refto{26}
$$
C_{JQ}={4\pi TSM\over 1-8\pi TM-4\pi ST^2}\sim {1\over T-T_c}
{}~,\eqno(31)
$$
where the critical temperature is given by
$$
T_c^{JQ}= {1\over 2\pi M}{1\over [3+\sqrt{3-q_{JQ}}]} ~.\eqno(32)
$$
and $q_{JQ}$ is given by the critical curve \eq{27}.

{}From the fundamental equations \(1), \(4) and \(8) we can obtain the
equation
of state of a Kerr black hole\refto{27}
$$
J={1\over 4\Omega^3}\left\{\Omega^2+{T^2\over 8}-{T\over 2}
\left[\Omega^2+
{T^2\over 16} \right]^{1/2}\right\} \left[\Omega^2+ {T^2\over 16}
\right]^{-1/2}~,\eqno(33)
$$
and Reissner - Nordtr\o m  black hole
$$
Q={1\over 4\pi T}(\Phi-\Phi^3)~.
$$
We have now all the elements to compute the critical exponents as
defined in
the section above. As we have already remarked, being $C_{J,Q}$ the
divergent
heat capacity, we write all quantities of interest in terms of the
Helmholtz
 free energy,
$F(T,\vec J)$. Thus, from the thermodynamical identities coming from
the
first law, eq \(3), we have
$$
K_{T}^{-1}=J{\partial^2 F\over\partial J^2}\biggr\vert_{T}~~,~~
\Omega=-{\partial F\over\partial J}\biggr\vert_{T}~~,~~
S=-{\partial F\over\partial T}\biggr\vert_{J}~~.\eqno(34)
$$
We can  obtain the  first two  critical exponents directly by
inspection
of eq \(31) and comparison with eq \(19):
$$
\alpha=1~~~,~~~\varphi=1~~.\eqno(35)
$$
Analogously, from eq \(30) (that diverges as $C_{J,Q}$) and
comparison with
eq \(21), we obtain
$$
\gamma=1~~~,~~~1-\delta^{-1}=1~\Rightarrow~\delta^{-1}\to0~~.\eqno(36
)
$$

We will now deal with the equations obtained from the first
derivatives of the Helmholtz potential. To obtain the corresponding
critical
exponents we choose a path either along a critical isotherm or at
constant
angular momentum
$J=J_c$ or constant charge $Q=Q_c$.
However, in this case the black hole equations of state just
reproduce the critical curves (such as eqs \(28)-\(29),
\(32) and others deduced from
them). In this case, we can formally assign a zero power  in eqs
\(20) and
\(22) corresponding to critical exponents:
$$
\beta\to0~~~,~~~\delta^{-1}\to0~~,
$$
$$
1-\alpha=0~~~,~~~\psi\to0~~.\eqno(37)
$$
One can easily check that the set of critical values given by eqs
\(35) - \(37)
satisfy the scaling laws, eqs \(24) (with $\beta\delta=1$).

One can also compute the so called gap exponents that describe the
fashion in
which the ratio of two successive derivatives of the free energy
diverges
\refto{20}. For example,
$$
{\partial^{n+1} F\over\partial J^{n+1}}\biggr/{\partial^{n}
F\over\partial
J^{n}}\sim\varepsilon^{-\Delta _n}~~.
$$
We obtain that $\Delta _n= 1$, independent of $n$.

The critical exponents can be obtained by noticing that the Helmholtz
potential, $F(T-T_c,J-J_c)-F_c $, close to criticality scales
linearly with
respect to its variables (as can be seen from eqs \(34) and the fact
that $S$
 and $\Omega$ are continuous functions at the phase transition).
Thus, $a_{\varepsilon}=1=a_J$, and eqs \(24'') produce the same
\(35)-\(37)
values for the critical exponents.

 Other five heat capacities can be computed, of which $C_{\Omega,Q}$
and
$C_{J, \Phi}$ exhibit also a singular behavior. The remaining
$C_{\Phi,Q}=
C_{J,\Omega}$ and $C_{\Omega,\Phi}$ being regular functions in the
allowed
set of values of the parameters\refto{26}.

A similar analysis to the previous case can be made for the other two
diverging heat capacities. In fact, the heat capacity at $\Omega$ and
$Q$
fixed can be written as
$$
C_{\Omega Q}=T{\partial S\over\partial T}\biggr\vert_{\Omega Q}=
-T{\partial^2 H_1\over\partial T^2}\biggr\vert_{\Omega Q}~, \eqno(c1)
$$
with the appropriate thermodynamics potential given by
$$
H_1=M-\vec\Omega\cdot\vec J-T S~.
$$
By use of eqs \(8) and \(1), we have \refto{26}
$$
C_{\Omega Q}={4S^3T\Phi^2\over \pi Q^2(2\Phi Q-M)}\sim {1\over
(T-T_c)}~,
\eqno(c2)
$$
where, in this case, the critical temperature is given by
$$
T_c^{\Omega Q}={1\over 4\pi M}\left({3q_{\Omega Q}-2\over
q_{\Omega Q}^2}\right)~, \eqno(c3)
$$
and the critical curve
$$
(1-q_{\Omega Q})^3-j_{\Omega Q}(1-q_{\Omega Q})^2-
({3\over2}q_{\Omega Q}-1)^2=0~,\eqno(c4)
$$
fixes the values of $q_{\Omega Q}$ and $j_{\Omega Q}$ in the range
$$
0\leq j_{\Omega Q}\leq 1/2~~; ~~~2/3\leq q_{\Omega Q}\leq 3/4~.
\eqno(c5)
$$
The two associated isothermal compressibilities are
$$
K_{T\Omega}^{-1}=Q{\partial \Phi\over\partial
Q}\biggr\vert_{T\Omega}=
Q{\partial^2 H_1\over\partial Q^2}\biggr\vert_{T\Omega}~, \eqno(c6)
$$
that  diverges as $C_{\Omega Q}$, and
$$
K_{TQ}={1\over J}{\partial J\over\partial \Omega}\biggr\vert_{TQ}=
{1\over J}{\partial^2 H_1\over\partial \Omega^2}\biggr\vert_{TQ}~,
\eqno(c6)
$$
that also diverges as $C_{\Omega Q}$ on the singular curve \eq{c4}.

We observe that the study of the critical transition suffered by the
black
hole when we keep $\Omega$ and $Q$ fixed, is governed by
exactly the same critical exponents as in the case of
the critical transition at $J$ and $Q$ fixed eqs \(35)-\(37).
Note, however, that the critical curve described by eq \(c4) is
different from that described by eq \(30) (see also Figure 1).

The remaining divergent heat capacity, $C_{J\Phi}$, can be
studied in a similar way:
$$
C_{J\Phi}=T{\partial S\over\partial T}\biggr\vert_{J\Phi}=
-T{\partial^2 H_2\over\partial T^2}\biggr\vert_{J\Phi}~, \eqno(c8)
$$
with
$$
H_2=M-\Phi Q-TS~,
$$
the appropriate thermodynamical potential for this case.

Again, use of eqs \(8) and \(1) produce \refto{26}
$$
C_{J\Phi }\sim {1\over T-T_c}~.
$$
The corresponding critical temperature is given by
$$
T_c^{J\Phi }={1\over 2\pi M}{K\over [j_{J\Phi }+
(K+1)^2]}~~;~~~K^2=1-j_{J\Phi }-q_{J\Phi }~.
\eqno(c9)
$$
The critical curve for this case can be written as
$$
-K^4+(j_{J\Phi }-3)K^3-(j_{J\Phi }+3)K^2+(j_{J\Phi }^2+4j_{J\Phi }-1)
K+6j_{J\Phi }-2j_{J\Phi }^2=0~,
\eqno(c10)
$$
for $q_{J\Phi }$ and $j_{J\Phi }$ in the range
$$
0\leq j_{J \Phi }\leq 2\sqrt{3}-3~~; ~~~0\leq q_{J\Phi }\leq 1~.
\eqno(c11)
$$
The two isothermal compressibilities of this case are
$$
K_{TJ}={1\over Q}{\partial Q\over\partial \Phi}\biggr\vert_{TJ}=
{1\over Q}{\partial^2 H_2\over\partial \Phi^2}\biggr\vert_{TJ}~~;
{}~~~K_{T \Phi }^{-1}=J{\partial \Omega\over\partial
J}\biggr\vert_{T\Phi}=
J{\partial^2 H_2\over\partial J^2}\biggr\vert_{T\Phi}~. \eqno(c12)
$$
Both of which diverges as $C_{J\Phi}$ on the curve  given by
eq\(c10).

We observe that again the critical exponents deduced from the
divergence
of the heat capacity and thermal compressibilities at $J$ and $\Phi$
fixed are those given by eqs \(35)-\(37). This result can in fact be
understood as a realization of the {\it hypothesis  of Universality}.
As can be seen from Figure 1, the critical curves for the three cases
studied are different, but
the critical exponents, according to the above mentioned hypothesis,
are
the same within each class as specified after eq \(24).
We also observe that the equality between the primed ($T\to T_c^-$)
and unprimed ($T\to T_c^+$) critical exponents
is trivially verified in each one of the three
transitions studied.

The values we have found for the critical exponents, (Eqs. \(35) -
\(37)),
correspond to those of the Gaussian model in two
dimensions\refto{51,52}.
It can be shown that this analogy also holds for the critical
exponents
derived from the correlation function. The implications of this
results
and the derivation of an effective Hamiltonian to describe black
holes
near criticality is presently under study and will be given in a
forthcoming
paper \refto{37}.

\head{4. Discussion}

We have reviewed how black holes follow  the four laws of
thermodynamics. We
then have seen how critical point transitions and critical exponents
can be
defined for fluid-gas and magnetic systems. It is well known that
critical
exponents fulfill the scaling laws or what we call the  fourth law of
thermodynamics. The scaling behavior in critical phenomena applies
to a vast
variety of systems. This lead us to postulate that the scaling
property also
holds for black holes. We have shown that in all the three cases of
phase
transitions we studied, this scaling laws are satisfied with
exponents
given by eqs \(35)-\(37).

It was argued \refto{31} that the second order phase transition
discovered by
Davies is of purely geometric origin and that the internal state of
the black
hole remains unaffected after the phase transition.
However, when Davies'
transition  is seen as a critical phase transition, the lack of
qualitative
change in the properties of the black hole can be understood as in
analogy
to what happens in the case of a liquid-vapor system; where near
criticality
no qualitative distinction can be made between phases. Note that in
this
case there is not such thing as a latent heat
\refto{32}(since $M$ remains continuous throughout the transition),
as it happens in magnetic critical transitions.
Besides, the critical transitions occur when we cold down the black
hole
with respect to the corresponding Schwarzschild temperature,
$T_S=1/(8\pi M)$,
by increasing its charge or angular momentum
at fixed total mass (see Figure 2).
Further, we have seen how black holes fulfil the scaling laws and
universality hypothesis, both characteristics of critical phase
transitions.

In fact, it was remarked by Hut \refto{32} that although this phase
transition
does not affect the internal state of the system it is physically
important as
it indicates  the transition from a region ($C_{JQ}<0$) where only a
microcanonical ensemble is appropriate (stable equilibrium if the
system is
isolated from the outside world) to a region ($C_{JQ}>0$) where a
canonical
ensemble can be also used (stable equilibrium with an infinite heat
bath).

Notably, scaling during phase transitions involving gravitational
effects
can be also found in
several scenarios, such as: Inflation at late times \refto{33,34},
the
strong field collapse of a massless scalar field coupled to gravity
\refto{35}
and in cosmic string networks\refto{36}.

The scaling hypothesis can also be extended to the static correlation
functions and their dynamical behavior \refto{19}. Interestingly
enough it was
noted \refto{38} that some of the correlation functions diverges for
extreme
Kerr \refto{39} and Reissner-Nordstr\o m \refto{40} black holes as
expected
when a phase transition takes place.
The situation with black hole thermodynamics seems to be that of
ordinary
thermodynamics before to the discovery of the underlying laws arising
from
statistical mechanics. In our case, we have to discover the quantum
theory
of gravitation to fully understand black holes. Nevertheless, we hope
the
lines developed in this work will help to go one step further towards
the
understanding of the deep connection between gravitation, quantum
theory and
statistical physics. In this sense, the relation among the
renormalization
group, conformal theory and critical phenomena,
the resemblance of the critical exponents
\(35)-\(37) and the Gaussian model in two dimensions\refto{51,52},
and the
explanation of  the broken ergodicity in black hole thermodynamics
\refto{8}
which in critical phenomena arises naturally \refto{41} are
interesting
enough hints to follow.

Indeed further research on this subject is needed to find the
underlying theory
behind all these analogies. We will deal on these and some other
subjects in a
forthcoming paper to be published elsewhere \refto{37}.

\vskip 12pt
\noindent
{\it Acknowledgements}

\noindent
\mangos

\vfill\eject

\input epsf

\centerline{\epsfxsize=10cm
\epsffile{ fig1.eps}}

\bigskip\bigskip\bigskip
{\bf Figure 1:} This figure shows the different critical curves in
the
$j~-~q$ plane (see \eqs{26}).
Black holes solutions remain within the triangle formed
by the Schwarzschild, extreme Reissner-Nordstr\o m and Kerr holes.
Heat
capacities are negative in the region closer to the origin
(Schwarzschild)
and positives in the opposite region, closer to the extreme Kerr -
Newman hole. Points labeled as $A$,
$B$, $C$, $D$, and $E$ have coordinates $(q,j)$ given by
$A=(0,3\sqrt{3}-3)$;
$B=(3/4,0)$; $C=(.73,.13)$; $D=(2/3,1/3)$ and $E=(1,0)$.

\vfill\eject

\centerline{\epsfxsize=11cm
\epsffile{ fig2.eps}}

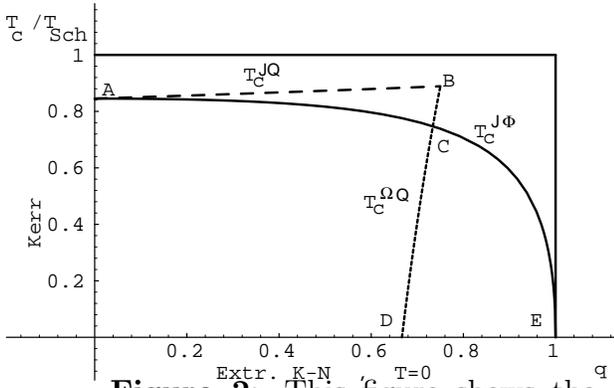
{\bf Figure 2:} This figure shows the critical temperatures
(normalized to
that of a  Schwarzschild hole of the same mass), as a function of the
critical
charge parameter $q$ (Projection onto the $T_c$ - $j$ plane gives
qualitatively
the same picture). Points $A$ and $B$ where curves meet represent the
Kerr and
Reissner-Nordstr\o m black holes respectively. At $A$,
$C_{J\Phi}=C_{JQ}
\equiv C_{J}$ and at $B$, $C_{\Omega Q}=C_{JQ}\equiv C_{Q}$ . Point
$C$,
gives the highest temperature for which all
three heat capacities are positive. Points $A$, $B$ and $C$ can be
classified
as bicritical points; while $D$ and $E$ can be classified as
tricritical
points\refto{50},(since there critical curves meet the first order
phase
transition curve $T=0$, which represents extreme holes and separates
black
holes from naked singularities\refto{38}).

\vfill\eject
\references

\refis{1} J. M. Bardeen, B. Carter and S. W. Hawking, {\it Commun.
Math.
Phys.}, {\bf 31}, 161 (1973).\par

\refis{2}  S. W. Hawking,  {\it Nature.}, {\bf 248}, 30 (1974); S. W.
Hawking,
{\it Commun. Math. Phys.} {\bf 43}, 199 (1975).\par

\refis{3}  G. W. Gibbons and  S. W. Hawking, {\it Phys. Rev.}, {\bf
D15}, 2738
(1977).\par

\refis{4} P. C. W. Davies, {\it Proc. R. Soc. Lond.}, {\bf A353}, 499
 (1977).\par

\refis{5} L. Smarr, {\it Phys. Rev. Lett.}, {\bf 30}, 71 (1973).\par

\refis{6} P. T. Landsberg, in {\it Black Hole Physics},   V. De
Sabbata and
Z. Zhang Eds, Nato Asi series, Vol. 364 (1992),  p 99-146.\par

\refis{7} P. T. Landsberg,  {\it Thermodynamics and Statistical
Mechanics},
Oxford University Press, (1978); New York: Dover (1991).\par

\refis{8} R. M. Wald, in {\it Black Hole Physics},  V. De Sabbata and
Z. Zhang
 Eds, Nato Asi series, Vol. 364 (1992),  p 55-97. \par

\refis{9} J. D. Bekenstein,  {\it Phys. Rev.}, {\bf D7}, 949 (1973);
ibid,
{\bf D7}, 2333 (1973); ibid, {\bf D9}, 3292 (1974).\par

\refis{10} S. W. Hawking, {\it Phys. Rev.}, {\bf D13}, 191  (1976).
\par

\refis{11}  S. W. Hawking, {\it Commun. Math. Phys.}, {\bf 25}, 152
(1972).\par

\refis{12} W. G. Unruh and R. M. Wald, {\it Phys. Rev.}, {\bf D25},
942
(1982).\par

\refis{13a} W. Israel, in {\it Black Hole Physics},   V. De Sabbata
and Z.
 Zhang  Eds, Nato Asi series, Vol. 364 (1992),  p 147-183.\par

\refis{13b} W. Israel, {\it Phys. Rev. Lett.}, {\bf 57}, 397
(1986).\par

\refis{14}  P. C. W. Davies, {\it Class. Quantum Grav.}, {\bf 6},
1909
 (1989).\par

\refis{15}  T. Andrews, {\it Phil. Trans. R. Soc.}, {\bf 159}, 575
(1869).\par

\refis{16} L. D. Landau and E. M. Lifshitz,  {\it  Statistical
Physics},
Pergamon Press, Oxford (1970).\par

\refis{17} M. F. Collins,  {\it  Magnetic Critical Scattering},
Oxford Univ.
 Press, Oxford (1989).\par

\refis{18} K. G. Wilson, {\it  Rev. Mod.  Physics.}, {\bf 55}, 583
(1983).\par

\refis{19} H. E. Stanley,  {\it  Introduction to Phase Transitions
and Critical
 Phenomena}, Oxford Univ. Press, Oxford (1971).\par

\refis{20} A. Hankey and H. E. Stanley, {\it Phys. Rev.}, {\bf B6},
3515
 (1972).\par

\refis{21} C. Garc\'\i a - Canal,  {\it  Apuntes de termodin\'amica},
La Plata report  (1985), unpublished.\par

\refis{22} B. Widom, {\it J. Chem. Phys.}, {\bf 43}, 3892 (1965); C.
Domb. and
D. L. Hunter, {\it Proc. Phys. Soc.}, {\bf 86}, 1147 (1965); L. P.
Kadanoff,
{\it Physics.}, {\bf 2}, 263 (1966) (see also refs. 1-32 of
\ref{20}).\par

\refis{23} K. G. Wilson and J. Kogut, {\it Phys. Rep.}, {\bf C12}, 75
(1974).\par

\refis{24} M. E. Fisher, in {\it Critical Phenomena}, F. J. W. Hahne
Ed,
Springer Verlag, Berlin (1982), p 1-139.\par

\refis{25} G. A. Baker,  {\it  Quantitative Theory of Critical
Phenomena},
Academic Press, Boston (1990).\par

\refis{26} D. Tranah and  P. T. Landsberg, {\it Collective
Phenomena},
{\bf 3}, 81  (1980).\par

\refis{27} I. O. Kamoto and O. Kaburaki,  {\it Mon. N. R. Astr. S.},
 {\bf 247}, 244, (1990).\par

\refis{28} P. T. Landsberg, {\it ``Thermodynamics" }, Interscience
Pub.
 Inc., N. Y. (1961). \par

\refis{29} P. T. Landsberg and D. Tranah, {\it Collective Phenomena},
{\bf 3}, 73, (1980).\par

\refis{30} A. Aharony and M. E. Fisher, {\it Phys. Rev.}, {\bf B27},
4394
  (1983). \par

\refis{31} L. M. Sokolowski and P. Mazur, {\it J.  Phys.}, {\bf A13
},
 1113  (1980).\par

\refis{32} P. Hut, {\it Mon. N. R. Astr. S.}, {\bf 180}, 379
(1977).\par

\refis{33} S. Matarrese, A. Ortolan and F. Lucchin, {\it Phys. Rev.
},
{\bf  D40}, 290 (1989).\par

\refis{34}  B. L. Hu, Class. Q. Gravity, {\bf 10}, S93  (1993).\par

\refis{35}  M. W. Choptuik, {\it Phys. Rev. Letters}, {\bf 70}, 9
(1992).\par

\refis{36}  T. W. B. Kibble, {\it Phys. Rev.}, {\bf D45 }, 1000
(1992).\par

\refis{37} C. O. Lousto, in preparation  (1993).\par

\refis{38}  A. Curir, {\it  Gen. Rel. Grav.}, {\bf 13}, 417
(1981).\par

\refis{39}  D. Pav\'on and J. M. Rub\'\i  , {\it Phys. Rev.}, {\bf
D37 },
 2052 (1988).\par

\refis{40}  D. Pav\'on and J. M. Rub\'\i  , {\it Phys. Rev.}, {\bf
D43 },
 2495 (1991).\par

\refis{41}   O. G. Mouristein,   {\it ``Computer Studies of Phase
Transitions
 and Critical Phenomena"}, F. J. W. Hahne Ed,
 Springer Verlag, Berlin (1984).\par

\refis{page} D. N. Page, in {\it Black Hole Physics},   V. De Sabbata
and Z.
 Zhang Eds, Nato Asi series, Vol. 364 (1992),  p 185-224.\par

\refis{Roman}  T. A. Roman, {\it  Gen. Rel. Grav.}, {\bf 20}, 359
(1988).\par

\refis{45} J. D. Bekenstein, {\it Phys. Rev.}, {\bf D12 }, 3077
(1975).\par

\refis{50} R. B. Griffiths, Phys. Rev. Lett., {\bf 24}, 715
(1970).\par

\refis{51} T. H. Berlin and M. Kac, Phys. Rev., {\bf 86}, 821
(1952).\par

\refis{52} A. Aharony, in {\it Critical Phenomena}, F. J. W. Hahne
Ed,
Springer Verlag, Berlin (1982), p 209-258.\par

\refis{ch} D. Christodolou, Phys. Rev. Lett., {\bf 25}, 1596
(1970).\par

\refis{cr} D. Christodolou and R. Ruffini, {\it Phys. Rev.},
{\bf D4}, 33552 (1971).\par

\endreferences
\end